\documentclass[
aps,
pra,
10pt,
showpacs,
tightenlines,
notitlepage,
longbibliography,
nofootinbib,
superscriptaddress,
twocolumn,
floatfix
]{revtex4-2}

\usepackage[T1]{fontenc}
\usepackage[english]{babel}

\usepackage{amsmath}
\usepackage{amssymb}
\usepackage{amsfonts}
\usepackage{mathtools}
\usepackage{bbold}
\usepackage{booktabs}

\usepackage{graphicx}
\usepackage{xcolor}

\usepackage{placeins}

\usepackage{algorithm}
\usepackage{algorithmic}

\usepackage{braket}
\usepackage{quantikz}

\usepackage{tikz}
\usepackage{tcolorbox}

\usepackage{microtype}

\usepackage[normalem]{ulem}

\usepackage[columnwise]{lineno} %

\usepackage[colorlinks=true,
linkcolor=blue,
citecolor=blue,
urlcolor=blue]{hyperref}

\begin{document}
\title{Efficient Quantum Algorithm for the Generalised Geography Game}

\author{Mriganka Sandilya}
\email{mrigankasandilya@gmail.com}

\affiliation{Center for Quantum Engineering, Research and Education (CQuERE), TCG CREST, Salt Lake, Sector 5, Kolkata 700091, India.}

\author{Shibdas Roy}
\email{roy.shibdas@gmail.com}
\affiliation{Center for Quantum Engineering, Research and Education (CQuERE), TCG CREST, Salt Lake, Sector 5, Kolkata 700091, India.}
\affiliation{Academy of Scientific and Innovative Research (AcSIR), Ghaziabad 201002, India.}

\begin{abstract}
The Generalized Geography Game ($GGG$) is a two-player combinatorial game played on a directed graph, where players alternately move a token along edges without revisiting previously visited vertices. In the existing literature, determining whether an arbitrary position in a graph is a winning position for either player is a well-known $PSPACE$-complete decision problem. Certain formulations of $GGG$ have been shown to fall outside the $PSPACE$-complete complexity class. While $PSPACE$-complete problems are believed to take exponential time for classical computation, quantum computation offers new possibilities. In this work, we present a bounded-error quantum polynomial-time ($BQP$) algorithm for solving the $GGG$ problem. The proposed algorithm runs in time $O(N^{2}/\epsilon^{6})$, which is polynomial in the input size $N$, and achieves bounded error with $\epsilon = O(1/\mathrm{poly}(N))$. This result demonstrates the potential of quantum algorithms to efficiently solve combinatorial game problems that are traditionally considered computationally hard in the classical setting.
\end{abstract}
\maketitle
\section{Introduction}
The complexity class $PSPACE$ consists of decision problems that can be solved by a classical Turing machine using an amount of memory that grows polynomially with the input size and possibly requiring an exponential number of steps \cite{arora2009computational, lichtenstein1980go, nielsen2010quantum, aaronson2008limitsof}. The complexity class $BQP$ (Bounded-Error Quantum Polynomial Time) consists of computational problems that can be solved by a quantum computer in polynomial time while allowing a bounded probability of error and cannot exceed the $PSPACE$ class. $PSPACE$-complete problems represent the most computationally difficult problems in $PSPACE$ class \cite{aaronson2008limitsof,nielsen2010quantum}. A prominent example of a $PSPACE$-complete combinatorial problem is the Generalized Geography Game (GGG) \cite{schaefer1978complexity, lichtenstein1980go, fox2022computational}. Combinatorial games are two-person, perfect-information games with no hidden information and no chance element, in which players move alternately according to fixed rules and the game must end after finitely many moves \cite{siegel2023combinatorial,fraenkel2012combinatorial, dorbec:hal-01429072, monti2018variants, burke2022quantum, fox2022computational}. In the existing literature, GGG is a two-player combinatorial game played on a directed graph \cite{lichtenstein1980go, storer1983complexity, FRAENKEL1993197, darmann2014shortest, iwata1994othello, rin2024arimaa, burke2022quantum, fox2022computational}, but can be extended to any number of players. In this game, players alternately move a token or a marker along directed edges and delete the vertex from which it is moved, with the game terminating when a player has no legal move left \cite{schaefer1978complexity, fraenkel1993undirected, bodlaender1993complexity, lichtenstein1980go, storer1983complexity, FRAENKEL1993197, iwata1994othello, darmann2014shortest, rin2024arimaa}.
The game is inspired by a well-known children’s word game in which players take turns naming cities,  the first letter
of the next city must be the same as the last letter of the previous city, and the next city
must not have been named before \cite{rin2024arimaa}. Determining whether an arbitrary position in a graph
is a winning position for either player is a well-known PSPACE-complete decision problem \cite{schaefer1978complexity, lichtenstein1980go, bodlaender1993complexity}.
\par {As an example, we consider two players $Alice$ and $Bob$, and we consider a finite directed graph generated from a set $S = \{\text{Agra, Agartala, Ambala, Assam}\}$ of $N = 4$ cities as the nodes. Suppose the players follow the sequence: $(Agra)_{\textit{Alice}} \rightarrow (Ambala)_{\textit{Bob}} \rightarrow (Assam)_{\textit{Alice}}$. In this case, $Alice$ initiates the game by choosing Agra, followed by $Bob$ selecting Ambala. $Alice$ then chooses Assam, successfully completing a favorable path, and therefore wins the game. Alternatively, consider the players follow the alternate sequence $(Agra)_{\textit{Alice}} \rightarrow (Ambala)_{\textit{Bob}} \rightarrow (Agartala)_{\textit{Alice}} \rightarrow (Assam)_{\textit{Bob}}$. In this scenario, $Alice$ diverts to Agartala after Ambala, allowing $Bob$ to capture Assam. As a result, $Bob$ gains the advantage and wins. This example illustrates how the sequence of moves and the strategic allocation of choices can directly influence the game's outcome, thereby determining the winning strategies for Alice and Bob.}
\par In the $PSPACE$ complexity class, GGG could serve as a benchmark for studying the complexity of combinatorial games and other $PSPACE$-complete problems. Computational complexity provides a framework for quantifying the difficulty of the problem in both time and space \cite{nielsen2010quantum, arora2009computational}. Within the classical setting, the GGG has been studied extensively under many formulations. These formulations, together with their combinations, differ along several dimensions, and most are $PSPACE$-complete. Key distinctions in the formulation includes the type of underlying graph is directed or undirected \cite{FRAENKEL1993197,fraenkel1993undirected,schaefer1978complexity,monti2018variants, fox2022computational}; structural constraints such as acyclicity, bipartiteness, planarity, and bounded degree of the vertex and edges and its out degree  \cite{bodlaender1993complexity, FRAENKEL1993197,lichtenstein1980go,fraenkel1993undirected, iwata1994othello, fox2022computational}; and the move model employed in the play, vertex-based or edge-based \cite{bodlaender1993complexity,lichtenstein1980go, FRAENKEL1993197, fox2022computational}. The formulation of each variant also depends on factors such as the removal of vertices or edges after a move \cite{fraenkel1993undirected, fox2022computational}, and the type of gameplay impartial, where both players share the same legal moves, or partisan games, where each player operates under a distinct set of allowable actions \cite{monti2018variants, FRAENKEL1993197, fraenkel1993undirected, fox2022computational}. Examples are \textit{n $\times$ n} chess \cite{storer1983complexity}, the shortest path problem \cite{darmann2014shortest}, Othelo \cite{iwata1994othello}, Nim \cite{burke2022quantum}, Go \cite{lichtenstein1980go}, Node Kayle \cite{schaefer1978complexity, burke2022quantum}, Arimma \cite{rin2024arimaa}, Generalized Hex \cite{schaefer1978complexity}, Shanon Switching Game \cite{even1976combinatorial}, Quantum Tic-Tac-Toe \cite{dorbec:hal-01429072}, avoid a cycle \cite{gates2024cycle} to mention a few.

\par Under certain circumstances, the complexity of GGG departs from $PSPACE$-completeness. Some formulations give polynomial-time and linear-time algorithms \cite{fraenkel1993undirected, arun2025geography, bodlaender1993complexity}. In some scenarios it also reduces to $NP$-hard \cite{FRAENKEL1993197, arun2025geography, fox2022computational} or extends to $EXP$-time \cite{arun2025geography}. 
Quantum concepts have inspired a new framework in combinatorial game theory \cite{burke2022quantum}. In this work, we present a quantum algorithmic framework to develop a gate-based Fault-Tolerant Quantum Computing (FTQC) algorithm that solves GGG in Bounded-error Quantum Polynomial (BQP) time. We consider the case where we take two players, $Alice$ and $Bob$, a directed graph and determine if there exists a possible winning strategy for the second player, $Bob$. The game starts at any arbitrary vertex of the graph.

We achieved this by using a quantum circuit involving controlled swap gates to yield all paths without repetition of vertices from the graph and Improved Quantum Phase Estimation (IQPE) \cite{harrow2009quantum}, and Density Matrix Exponentiation (DME) \cite{kjaergaard2022demonstration,  harrow2009quantum} approach to determine the valid path to traverse in the graph. The algorithm complexity is $O(N^2/ \epsilon^6)$ and the overall error probability of the algorithm is less than or equal to ${1/3}$.  
\section{Method}
\begin{figure}[htbp]
    \flushright
    \includegraphics[width=1.1\linewidth]{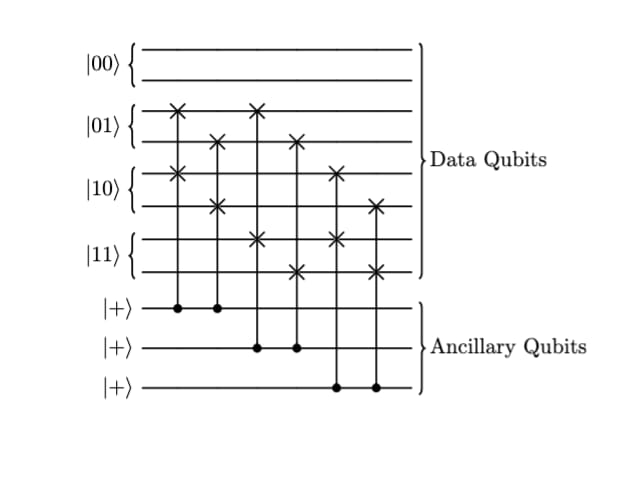}
    \caption{Quantum circuit using controlled swap gates to create a superposition of all permutations with starting city $\ket{0} = \ket{00}$ for $N = 4$.}
    \label{fig:ckt}
\end{figure}
We are given a finite, directed graph generated from a set of $N$ cities, where each city corresponds to a vertex. For example, we consider the graph having $4$ vertices and we use $\log M:= \lceil \log N \rceil = 2 $ qubits to index the cities in the vertices using 4 levels:  $\ket{0} = \ket{00}, \ket{1} = \ket{01}, \ket{2} = \ket{10}, \ket{3} = \ket{11}$. The edges connecting the vertices of the graph from city $a$ to city $b$ are encoded using phase factors $\delta_{ab}$, in our example, $a,b \in \{1,2,3,4\}$. For every valid edge in the graph between $a$ to $b$, we assign $\delta_{ab} = 1,$ $\forall$ $a,\, b < N, a \neq b$. For the edges that do not exist, we assign $\delta_{ab} = 2, \forall \hspace{2pt} a,\,b < N$ and $\delta_{ab} = 0 \hspace{0.1cm} \forall \hspace{0.11cm} a = b$. In addition, prior to executing the algorithm, we precompute the sum of the assigned phase values of all $2 \times {^N}C_2$ number of edges, and denote this quantity by $\phi_{\text{sum}} = \sum_{a,b}\delta_{ab}$. Each $\delta_{ab}$ is divided by $\phi_{\text{sum}}$ such that the normalized phase factors are $\phi_{ab} = \frac{\delta_{ab}}{\phi_{sum}}$. In the graph, the normalized phase factor value, $\phi_{ab}$ replaces the phase factors $\delta_{ab}$. We denote the normalized phase factor values for valid edges as $\phi^{V}_{ab}$.
\par In the graph, to determine the valid path to be traversed without repetition of any cities in the graph, which corresponds to a winning strategy for $Bob$, 

the graph formed by the given set of cities may or may not be fully connected. Our aim is to develop a quantum algorithm that would take polynomial time in $N$ to identify this path, if one exists, in the graph. 

\par Generally, to generate all the possible permutations of valid paths for $N$ number of qubit data registers, we consider $\lceil\log N\rceil$ number of qubits and initialise it to $\ket{1}, \ket{2}, \ket{3}, \ket{4}, \ldots \ket{N-1}$, respectively.  To achieve this, we apply, $^{(N-1)}C_{2}$ number of single-qubit ancilla registers, each initialized to state $\ket{+} = \frac{1}{\sqrt{2}}(\ket{0}) + \ket{1})$, and applying $\lceil \log N \rceil$ number of controlled swap gates on each of every combination of two of the $N$ qubit data registers with one ancilla register as the control qubit. During the process of adding the controlled swap, we do not perform the controlled swap operation on the vertex from which the game begins. In our example, $Alice$ starts the game with the city $\ket{00} = \ket{0}$. Please refer to the circuit in Fig.~\ref{fig:ckt}. Furthermore, this circuit is closely similar to Fig. 1 of Ref.~\cite{sharma2025efficient}, which constructs all permutations of cycles that include the maximum number of edges in the graph. 

\par Utilizing the normalized phase factors $\phi_{ab}$ (as mentioned earlier), we create a diagonal matrix $\Phi$ as:

{\small
\begin{equation}\label{eq:phi}
\begin{split}
\Phi &= \phi_{00}\ket{00}\bra{00}+\phi_{01}\ket{01}\bra{01}+\phi_{02}\ket{02}\bra{02}+\phi_{03}\ket{03}\bra{03}\\
&+\phi_{10}\ket{10}\bra{10}+\phi_{11}\ket{11}\bra{11}
+\phi_{12}\ket{12}\bra{12}
+\phi_{13}\ket{13}\bra{13}\\
&+\phi_{20}\ket{20}\bra{20}+\phi_{21}\ket{21}\bra{21}+\phi_{22}\ket{22}\bra{22}+\phi_{23}\ket{23}\bra{23}\\
&+\phi_{30}\ket{30}\bra{30}+\phi_{31}\ket{31}\bra{31}+\phi_{32}\ket{32}\bra{32}+\phi_{33}\ket{33}\bra{33},
\end{split}
\end{equation}
}
\\
Next, using the diagonal matrix $\Phi$ (Eq.\ref{eq:phi}) we create a unitary matrix $U$ such that it encodes all edges into the normalized phase factors $\phi_{ab}$ of their eigenstates: 
\begin{equation}
\label{eq : unitrary general}
U = \sum\limits_{a,b=0}^{N-1} e^{i\phi_{ab}}\ket{ab}\bra{ab}.
\end{equation}
In the rest of the paper, without any loss of generality, we will consider $M = N$. The diagonal unitary matrix $U$ for the vertices represented by index values $j = 0, 1, 2, 3$ and for $N = 4 \hspace{0.1cm} (= M)$ is:
{\footnotesize
\begin{align}\label{eq:unitary}
\begin{split}
U &= e^{i\phi_{00}}\ket{00}\bra{00} + e^{i\phi_{01}}\ket{01}\bra{01} + e^{i\phi_{02}}\ket{02}\bra{02} + e^{i\phi_{03}}\ket{03}\bra{03} \\
&+ e^{i\phi_{10}}\ket{10}\bra{10} + e^{i\phi_{11}}\ket{11}\bra{11} + e^{i\phi_{12}}\ket{12}\bra{12} + e^{i\phi_{13}}\ket{13}\bra{13} \\
&+ e^{i\phi_{20}}\ket{20}\bra{20} + e^{i\phi_{21}}\ket{21}\bra{21} + e^{i\phi_{22}}\ket{22}\bra{22} + e^{i\phi_{23}}\ket{23}\bra{23} \\
&+ e^{i\phi_{30}}\ket{30}\bra{30} + e^{i\phi_{31}}\ket{31}\bra{31} + e^{i\phi_{32}}\ket{32}\bra{32} + e^{i\phi_{33}}\ket{33}\bra{33}.
\end{split}
\end{align}
}
When simulating the unitary $U$ at Eq.~\ref{eq:unitary}, we perform exponentiation of the diagonal matrix $\Phi$ to yield $U = e^{i\Phi \tau_{\circ} }$, where $\tau_{\circ}$ is the time variable in simulating $U$. Moreover, $\Phi$ in Eq.~\ref{eq:phi}, being diagonal and sparse, $U$ can be simulated efficiently \cite{harrow2009quantum, berry2007efficient}. Moreover, we shall have $\tau_{\circ} = O (1/\epsilon)$ and $\epsilon/2$ is the estimation error in trace distance, and hence, the maximum probability of estimation error. We set $\epsilon = O(1/\text{poly}(N))$ so that the simulation time $\tau_{\circ}$ for $U$ scales in a polynomial manner rather than exponentially with $N$, ensuring that the total error probability of the algorithm remains below $1/3$.

\par
\begin{tcolorbox}[title = Box 1 : Improved Quantum Phase Estimation (IQPE).]
We start with an initial state $\ket{\Lambda_0} \ket{u_j}$, where $\ket{u_j}$ is the $j$-th eigenstate of the Hermitian matrix $\Gamma$, which we exponentiate. The state $\ket{\Lambda_0}$ is defined as:
$\ket{\Lambda_0} := \sqrt{\frac{2}{T}} \sum_{\iota=0}^{T-1} \pi\left(\iota + \frac{1}{2} \right) \sin\left( \frac{\pi(\iota + \frac{1}{2})}{T} \right) \ket{\iota}$ for some large $T$. The initial state $\ket{\Lambda_0}$ can be prepared up to some error $\varepsilon_\Lambda$ in time $\mathrm{poly}\left( \log_2(T / \varepsilon_\Lambda) \right)$ (see Section A of the Supplementary Material of Ref. \cite{harrow2009quantum}). We apply the conditional Hamiltonian evolution:
$\sum_{\iota=0}^{T-1} \ket{\iota}\bra{\iota} \otimes e^{i \Gamma \iota t_0 / T}$ on the initial state in both registers, and then apply the quantum Fourier transform (QFT) on the first register to obtain the state:
$\sum_{p=0}^{T-1} \nu_p^{(j)} \ket{p} \ket{u_j}$. Defining the estimate $\tilde{r}_p$ of the $p$-th eigenvalue $r_p$ of $\Gamma$ as:
$\tilde{r}_p := \frac{2\pi p}{t_0}$, we can relabel the Fourier basis states $\ket{p}$ to obtain:
$\sum_{p=0}^{T-1} \nu_p^{(j)} \ket{\tilde{r}_p} \ket{u_j}$. If the phase estimation is perfect, then $\nu_p^{(j)} = 1$ if $\tilde{r}_p = r_j$, and 0 otherwise. So, we obtain the state $\ket{\tilde{r}_j} \ket{u_j}$, from which we get the estimate of $r_j$ upon measuring the first register. This quantum phase estimation method errors by $\varepsilon = \mathcal{O}(1/t_0)$ in estimating $r_j$ \cite{harrow2009quantum}, where $\varepsilon/2$ is the error in trace distance (see just before Section A and just before theorem 6 in the supplementary Material of Ref. (\cite{harrow2009quantum}).
\label{Box 1 : Improved Quantum Phase Estimation}
\end{tcolorbox}
\par We create an $N$ qubit register state $\rho$ (Eq.\ref{eq : rho}) by following the circuit shown in Fig.~\ref{fig:ckt} and subsequently tracing out the ancilla registers. We have
\begin{equation}
    \label{eq : rho}
    \rho = \frac{1}{2^{^{^{(N-1)}C_{2}}}}\sum_{\mu}\beta^{2}_{\mu}\ket{\mu}\bra{\mu},
\end{equation}
where ${\sum_{\mu}\beta^{2}_{\mu} = 2^{^{(N-1)}C_{2}}}$.

\par
\begin{figure}[htbp]
\centering
    \includegraphics[scale = 0.23]{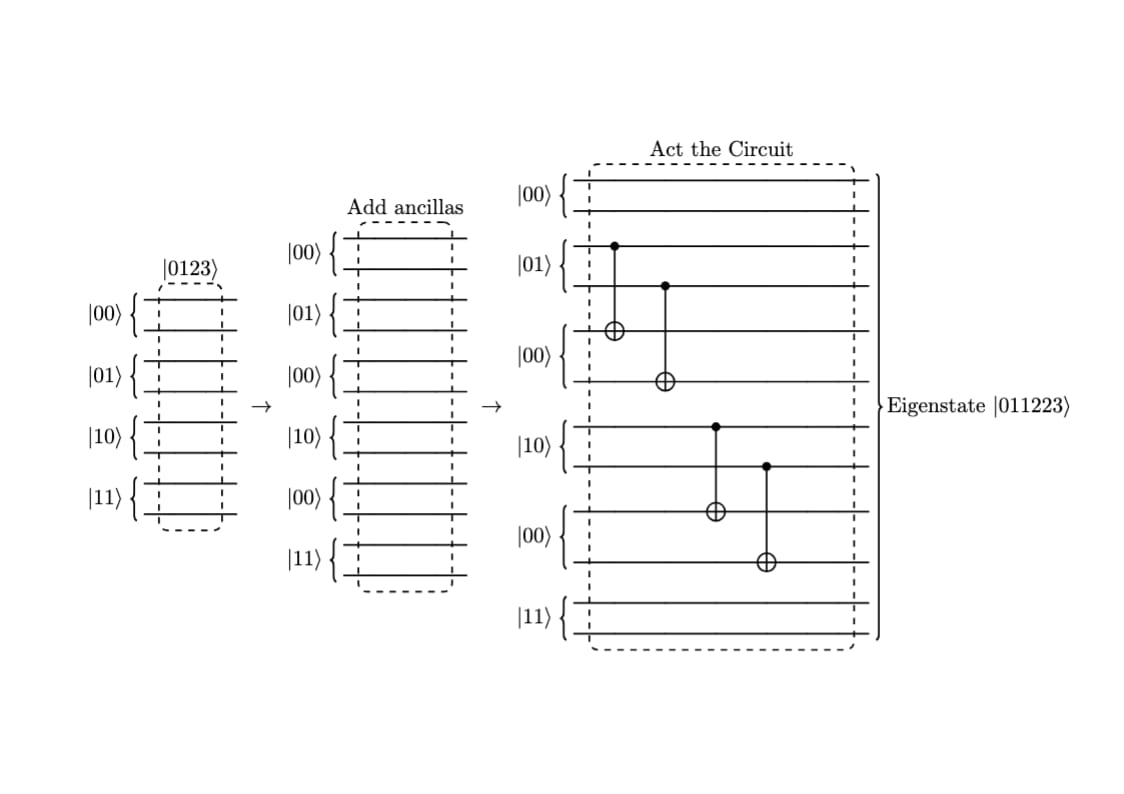}
    \caption{Illustration of a Quantum circuit to augment terms like $\ket{0123}$ in $\rho$ to $\ket{011223}$ in $\sigma$ for $N = 4$}
    \label{fig:register-state}
\end{figure}

 In the next step, we expand the state ${\rho}$ in Eq.~\ref{eq : rho} from an $N$ qubit register state to a new $2(N-1)$ qubit register state, forming a valid eigenstate of the unitary $V$ to perform IQPE (mentioned later). For example, we expand the state corresponding to a path traversing the maximum number of possible edges $\ket{0} \rightarrow \ket{1} \rightarrow \ket{2} \rightarrow \ket{3}$, which is $\ket{\mu} = \ket{0123}$ to $\ket{j} = \ket{011223}$. The eigenstate $\ket{j}$ corresponds to the longest path that can be obtained from a given graph. The quantum circuit required to attain this is shown in Fig.~\ref{fig:register-state}, constructed by adding $(N-2)$ ancillary qubits, each initialised in $\ket{0}$ and applying $CNOT$ operations at appropriate places to create copies of every register in $\rho$. We call this new state $\sigma$, with $\sum_{j}\beta^{2}_{j} = 2^{^{(N-1)}C_{2}}$ as,
\begin{equation}
    \sigma = \frac{1}{2^{^{(N-1)}C_{2}}}\sum_{j}\beta^{2}_{j}\ket{j}\bra{j}.
    \label{sigma-data-registers}
\end{equation}
{\sloppy
{\begin{figure}[htbp]
    \centering
    \includegraphics[scale=0.23]{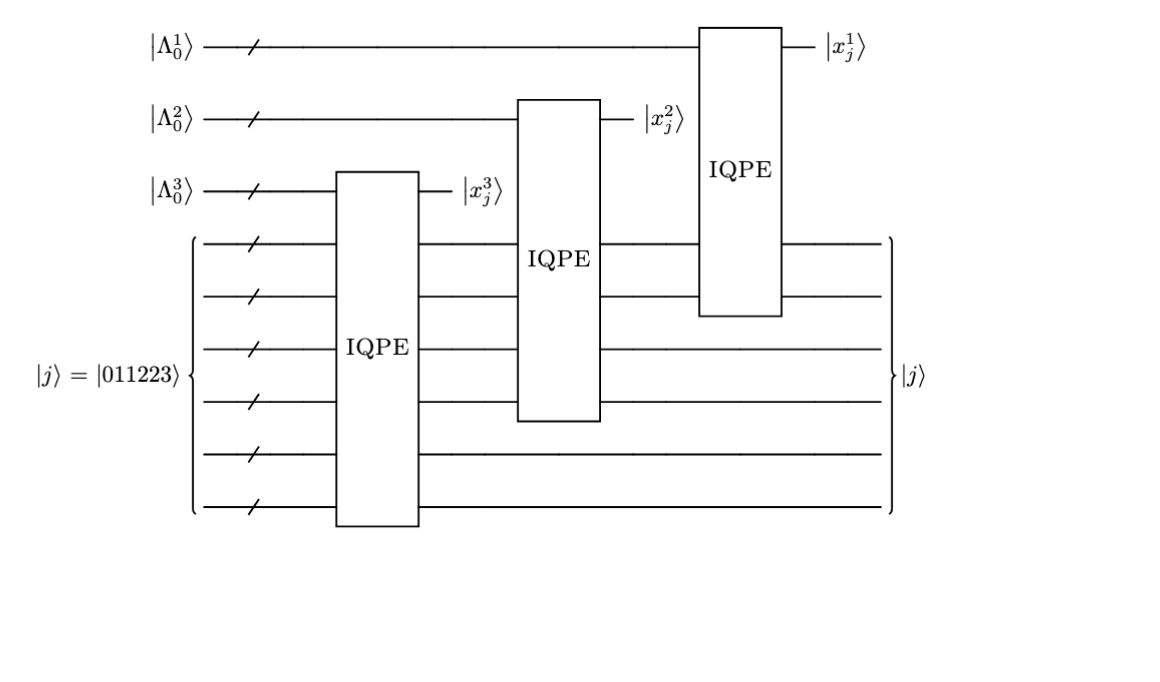}
    \caption{Quantum circuit representing the iterative Improved Quantum Phase Estimation Process for the eigenstate $\ket{j} = \ket{011223}$}
    \label{improved_quantum_phase_estimation circuit}
\end{figure}}
}

\par Quantum Phase Estimation (QPE) is an algorithm that efficiently computes the phase of a given eigenstate of a unitary in polynomial time, depending on the required precision of the phase estimate \cite{nielsen2010quantum}. In our algorithm, we use the Improved Quantum Phase Estimation (IQPE) method from Ref. \cite{harrow2009quantum}, which is outlined in \hyperref[Box 1 : Improved Quantum Phase Estimation]{Box 1}, in an iterative manner. To carry out the iterative procedure, we create $(N-1)$ copies of the $U$ to form the unitary $V = U^{\otimes l}$, where $l$ is initialized to $(N-1)$ and is decreased by one at each iteration. $l$ denotes the number of edges present in a eigenstate $\ket{j}$ and $(N-1)$ is the maximum number of edges of the path that can be generated from the given graph.
\par Now, we perform IQPE of $\sigma$ with the $V$. After each iteration of IQPE on $\ket{j}$ using $V$ in the iterative way, the number of edges in the eigenstate is reduced by one. Consequently, $l$ decreases by one and hence the dimension of $V$. This iterative process continues until $l = 1$. 

\par For example, as shown in  Fig.~\ref{improved_quantum_phase_estimation circuit}, the initial eigenstate in the qubit register is $\ket{j} = \ket{011223}$, encoding a path with the maximum number of edges for a given graph with $N = 4$ cities. It undergoes IQPE with the unitary $V = U^{\otimes 3}$, yielding the phase estimate $\ket{x_{011223}^{3}}$. In the next iteration, using the same qubit register and reducing one edge, IQPE is performed with the eigenstate $\ket{j} = \ket{0112}$ using unitary $V = U^{\otimes 2}$, resulting in the phase estimate $\ket{x^2_{0112}}$. Finally, in the last iteration, further reducing another edge, IQPE is applied to the eigenstate $\ket{j} = \ket{01}$ with the unitary $V = U^{\otimes 1}$, producing the phase estimate $\ket{x^1_{01}}$. 

This procedure yields the output state $\gamma$, where the phase estimate for $l$ iterations is expressed as $\ket{x_{j}^{l}} = \ket{\sum_{m=1}^{l} \phi_{ab}^{m}}$, with $\phi_{ab}^{m}$ denoting the sums of the predefined normalized phase factors $\phi_{ab}$ as:
\begin{equation} 
\label{eq : gamma}
    \gamma = \frac{1}{2^{^{(N-1)}C_{2}}}\sum_{j}\beta^{2}_{j}\Big(\bigotimes_{l=(N-1)}^{1}|x_{j}^{l}\rangle\langle x_{j}^{l}| \Big)\otimes \ket{j}\bra{j}.
\end{equation}
In Eq.~\ref{eq : gamma}, the state $\gamma$ encodes all the eigenstates incorporated within $\ket{j}$ with fixed starting vertex, together with the respective phase estimates $\ket{x_{j}^{l}}$. In order to determine the valid path, we add to $\gamma$ an ancilla register of $n$ qubits for each $l$ in the state $\ket{0^{\otimes n}}$ and thereafter rotating this state conditioned on $\ket{x_{j}^l}$. We call this state $\Omega$ : 
\begin{equation}
\begin{aligned}
\label{eq : control-rotation}
\Omega = \frac{1}{2^{^{(N-1)}C_{2}}}\sum_{j}\beta_{j}^2 \Big(\bigotimes_{l=(N-1)}^{1} \ket{x_{j}^{l}} \bra{x_{j}^{l}}\Big) \otimes \ket{j}\bra{j} \\ \Bigg\{ \bigotimes_{l=(N-1)}^{1}
\left[ \left(\sqrt{1 - \frac{(x_{j}^{l})^2}{(E_{l})^2}} \ket{0^{\otimes n}} +  \frac{x_{j}^{l}}{E_{l}} \ket{1^{\otimes n}}\right)\right.\\
\left.\left( \sqrt{1 - \frac{(x_{j}^{l})^2}{(E_{l})^2}} \langle 0^{\otimes n}| +  \frac{x_{j}^{l}}{E_{l}} \langle 1^{\otimes n}|\right)\right] \Bigg\}.
\end{aligned}
\end{equation}
In Eq.~\ref{eq : control-rotation}, the value of $E_{l} = l \times{(\phi^{V}_{ab})}$, which is the product of the number of edges in an eigenstate $\ket{j}$ times the normalised phase factor value of the valid edges of the given graph. For example for $\ket{j} = \ket{011223}$, the value of $E_{l=3}$ is  $3 \times \phi^{V}_{ab}$, for $\ket{j} = \ket{0112}$, $E_{l=2}$ is  $2 \times \phi^{V}_{ab}$ and for $\ket{j} =  \ket{01}$, $E_{l=1}$ is  $1 \times \phi^{V}_{ab}$.

\par The effective state of all the rotated single ancilla qubit for all $l=(N-1)$ to $1$, in general, taken together, is :
\begin{equation}
\begin{aligned}
\label{eq : effective_state}
    \chi := \frac{1}{2^{^{N-1}C_{2}}} \sum_{j}\beta_{j}^2\Bigg\{ \bigotimes_{l=N-1}^{1}\left[\left(1 - \frac{(x_{j}^{l})^2}{E_{l}^2}\right) \ket{0}\bra{0} +{} \right. & \\
    \left. \frac{(x_{j}^{l})^2}{E_{l}^2} \ket{1}\bra{1}\right] \Bigg\}.
\end{aligned}
\end{equation}

Next $\chi$ is exponentiated to obtain the unitary $Y = e^{i \chi \tau_{1}}$. To simulate $Y$ within an error $\epsilon$, we apply Eq.~\ref{eq : copies of ancilla} a total of 
$n = O\left({\tau_{1}^2}/{\epsilon}\right)$ 
times. In this process, $n$ copies of the ancillary states $\chi$ are generated from the corresponding $n$ copies of the ancillary registers \cite{lloyd2014quantum, darmann2014shortest}.
\begin{equation}
\label{eq : copies of ancilla}
\mathrm{Tr}_{\text{anc}}\left[e^{i S \Delta \tau_{1}} (\varrho \otimes \varsigma) e^{-i S \Delta \tau_{1}} \right]
= \varsigma - i \Delta \tau_{1} [\varrho, \varsigma] + \mathcal{O}(\Delta \tau^2_{1}).
\end{equation}
In Eq.\ref{eq : copies of ancilla}, $\tau_{1} = n\Delta\tau_{1}$, $\varsigma$ is another state and $S$ is the swap operator which is sparse and hence $e^{iS\Delta\tau_{1}}$ can be performed efficiently.
\begin{figure}[htbp]
    \centering
    \includegraphics[scale=0.55
    ]{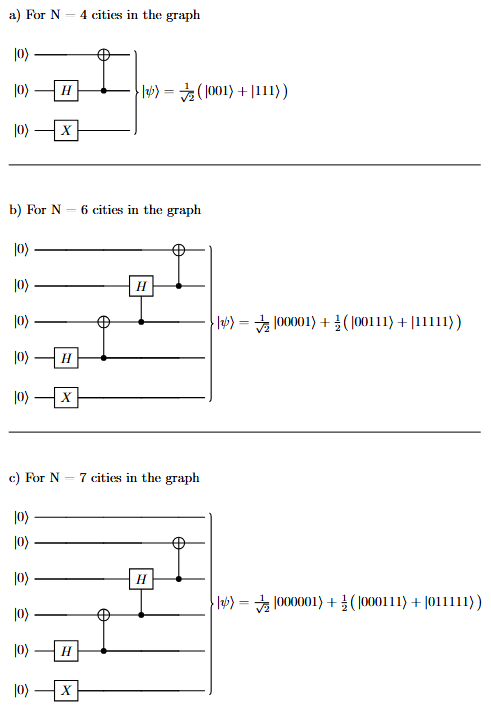}
    \caption{Quantum circuit producing required eigenstates for quantum phase estimation with $Y$, for different cities in the graph. a) For $N = 4$ b) $N = 6$ and c) $N = 7$ cities in the graph.}
    \label{fig : QPE with N = 4 cities with 3 edges}
\end{figure}
\par In the next stage, we perform IQPE on $Y$ using the eigenstates generated by the quantum circuit shown in Fig.~\ref{fig : QPE with N = 4 cities with 3 edges}a, denoted by $\ket{\psi}$, for $n^{\prime}$ iterations. This requires creating $n^{\prime}$ copies of $Y$ and the state $\ket{\psi}$; the reason for performing $n^{\prime}$ iterations is explained later. 
\par The quantum circuit shown in Fig.~\ref{fig : QPE with N = 4 cities with 3 edges} encodes a superposition of eigenstates, with each qubit rotated in proper order in either $\ket{0}$ or $\ket{1}$ to represent a set of valid edges traversed, which, in this case corresponds to a winning path for the second player i.e., $Bob$, for $N = 4,5 \hspace{2pt} \& \hspace{2pt} 6$ cities in the graph, for Fig.~\ref{fig : QPE with N = 4 cities with 3 edges}a, Fig.~\ref{fig : QPE with N = 4 cities with 3 edges}b, and Fig.~\ref{fig : QPE with N = 4 cities with 3 edges}c respectively. The number of qubits present in the eigenstates represents the number of edges in a particular path. In general, for $N$ cities, the quantum circuit follows a generic pattern. As the number of edges in a graph increases along with increase in the number of cities or vertices in a graph, it employs a combination of $X$-gate, $H$-gate, controlled rotation gates, controlled Hadamard gate, and additional qubits to generate the valid superposition of eigenstates. For $N$ cities, since there are $(N - 1)$ edges, we have $(N - 1)$ qubits in the state $\ket{\psi}$.\

After performing the IQPE of $Y$ with $\ket{\psi}$, we get the output $\ket{\zeta}$ with the phase estimate $\ket{\omega_k}$, $z_k$ being the probability amplitude and $k$ number of terms as:
\begin{equation}
\label{eq : zeta}
\ket{\zeta} = \sum_{k=1} ^ {\lfloor \frac{N}{2} \rfloor} (z_k) \Big[\ket{\omega_k} \otimes \ket{\psi_k} \Big].
\end{equation}
From $\ket{\zeta}$, using the effective state of the first register i.e., $\ket{\omega_k}$, we obtain the state $\Lambda$ as : 
\begin{equation}
    \label{lambda}
    \Lambda = \sum_{k=1}^{\lfloor\frac{N}{2}\rfloor} (z_k^2)\ket{\omega_k}\bra{\omega_k}.
\end{equation}
In Eq.~\ref{lambda}, $z_k^2$ is the associated probability.
Then we perform DME of the density matrix $\Lambda$ to obtain $W = e^{i\Lambda\tau_2}$. To simulate $W$ with an error $\epsilon$, we require $n^{\prime} = O(\tau_2^2/\epsilon)$ copies of $\Lambda$, obtained by repeating the earlier IQPE process of $Y$ with generated eigenstates $\ket{\psi}$ for $n^{\prime}$ number of times, resulting in $n^{\prime}$ copies of $\ket{\zeta}$.
\par Next we perform IQPE of $W$ with the eigenstate $\ket{0^{\otimes n_{1}}}$ $n_1 = \lceil\log_{_{2}}(dim(\Lambda))\rceil$, consisting of $m^{\prime}$ number of qubits in the phase estimate registers. If, upon measuring the phase–estimation register, the outcome is $\frac{1}{2\pi}$, there is no winning strategy for $Bob$. If the phase estimate is less than $\frac{1}{2\pi}$, then there exists a possible winning strategy for $Bob$.

\section{Algorithm}
\begin{enumerate}
    \item \label{algo_step 1} We are given two players, $Alice$ and $Bob$, and a finite directed graph representing a set of $N$ cities. The edges connecting the vertices of the graph from city $a$ to city $b$ are encoded using phase factors $\delta_{ab}$, in our example, $a,b \in \{1,2,3,4\}$. For every valid edge in the graph between $a$ to $b$, we assign $\delta_{ab} = 1,$ $\forall$ $a,\, b < N, a \neq b$. For the edges that do not exist, we assign $\delta_{ab} = 2, \forall \hspace{2pt} a,\,b < N$ and $\delta_{ab} = 0 \hspace{0.1cm} \forall \hspace{0.11cm} a = b$. We precompute the sum of the assigned phase values of all $2 \times {^N}C_2$ number of edges, and denote this quantity by $\phi_{\text{sum}} = \sum_{a,b}\delta_{ab}$. Each $\delta_{ab}$ is divided by $\phi_{\text{sum}}$ such that the normalized phase factors are $\phi_{ab} = \frac{\delta_{ab}}{\phi_{sum}}$, of a unitary $U = e^{i\Phi \tau_{\circ}}$, where  
    $\Phi = \sum_{a,b = 0}^{N-1}\phi_{ab}\ket{ab}\bra{ab}$. We denote the normalized phase factor values for valid edges as $\phi^{V}_{ab}$. We generate $(N-1)$ copies of $U$ for $V = U^{\otimes l}$, and the parameter $l$ is initialized to $(N-1)$ and is decreased by one at each iteration upon performing improved quantum phase estimation.  

    \item \label{algo_step 2} We create $\lceil \log N \rceil$ qubits to index the cities in the vertices and initialize each of the $N$ qubit data registers to $\ket{0}, \ket{1}, \ket{2}, \ldots, \ket{N-1},$ respectively. We generate all the permutations of the paths from $0,1,2, \ldots N-1$ without repetition of vertices by applying $^{(N-1)}C_2$ number of single-qubit ancilla registers, each of which is prepared to the state $\ket{+} = \frac{1}{\sqrt{2}}(\ket{0} + \ket{1})$ and then applying $\lceil \log N \rceil$ number of controlled swap gates on each of every combination of two of the $N$ qubit data registers where one ancilla register is the control qubit. During the process of adding the controlled swap, we do not perform the controlled swap operation on the vertex from which the game begins.
    
    \item \label{algo_step 3} We initialize a $N$ qubit register state $\rho = \frac{1}{2^{^{(N-1)}C_{2}}} \sum_{\mu}\beta_{u}^2\ket{\mu}\bra{\mu}$, where $\sum_{\mu}\beta_{\mu}^2 = 2^{^{(N-1)}C_{2}}$. Then we expand $\rho$ to a $2(N-1)$ qubit register state $\sigma = \frac{1}{2^{^{(N-1)}C_{2}}}\sum_{j}\beta_{j}^{2}\ket{j}\bra{j}$, where $\sum_{j}\beta_{j}^{2} = 2^{^{(N-1)}C_2}$. This is obtained by adding $(N-2)$ qubits, each initialised to state $\ket{0}$, and enacting it with $CNOT$ gates at appropriate places.
    
    \item \label{algo_step 4} We iteratively perform Improved Quantum Phase Estimation (IQPE) on the eigenstates $\ket{j}$ of $\sigma$, using the unitary operator $V = U^{\otimes l}$. For an eigenstate $\ket{j}$ after each iteration of IQPE, the number of edges in the eigenstate is reduced by one. Consequently, $l$ decreases by one and hence the dimension of $V$. The procedure yields the output state $\gamma$, where the phase estimate for $l$ iterations is expressed as $\ket{x_{j}^{l}} = \ket{\sum_{m=1}^{l} \phi_{ab}^{m}}$, with $\phi_{ab}^{m}$ denoting the sums of the predefined normalized phase factors $\phi_{ab}$, as:
    \begin{align*}  
    \gamma = \frac{1}{2^{^{(N-1)}C_{2}}}\sum_{j}\Big(\bigotimes_{l=(N-1)}^{1}\beta^{2}_{j}|x_{j}^{l}\rangle\langle x_{j}^{l}| \Big)\otimes \ket{j}\bra{j}.
    \end{align*}
    The state $\gamma$ encodes all the eigenstates incorporated within $\ket{j}$ with fixed starting vertex, together with the respective phase estimates $\ket{x_{j}^{l}}$.
    
    \item \label{algo_step 5} We add to $\gamma$, an ancilla register of $n$ qubits for each $l$ in the state $\ket{0^{\otimes n}}$ and thereafter rotating this state conditioned on $\ket{x_{j}^l}$. We call this state $\Omega$: 
    \begin{equation*}
    \begin{aligned}
    \label{eq : control-rotation algo}
    \Omega = \frac{1}{2^{^{(N-1)}C_{2}}}\sum_{j}\Big(\bigotimes_{l=(N-1)}^{1} \beta_{j}^2 \ket{x_{j}^{l}} \bra{x_{j}^{l}}\Big) \otimes \ket{j}\bra{j} \\ 
    \left[ \left(\sqrt{1 - \frac{(x_{j}^{l})^2}{(E_{l})^2}} \ket{0^{\otimes n}} +  \frac{x_{j}^{l}}{E_{l}} \ket{1^{\otimes n}}\right)\right.\\
    \left.\left( \sqrt{1 - \frac{(x_{j}^{l})^2}{(E_{l})^2}} \langle 0^{\otimes n}| +  \frac{x_{j}^{l}}{E_{l}} \langle 1^{\otimes n}|\right)\right],
    \end{aligned}
    \end{equation*}
    where $E_{l} = l \times{(\phi^{V}_{ab})}$, which is the product of the number of edges in an eigenstate $\ket{j}$ times the normalised phase factor value of the valid edges of the given graph.

    \item \label{algo_step 6} After rotation, we exponentiate the effective state $\chi$ of each ancilla qubit, using $n$ copies of $\chi$ from $n$ ancilla qubits, to obtain the unitary $Y:= e^{i\chi\tau_1}$. Then we perform IQPE of $Y$ with $\ket{\psi}$, generated by the quantum circuit shown in Fig.~\ref{fig : QPE with N = 4 cities with 3 edges}a, $n^{\prime}$ times. We get the output $\ket{\zeta}$ with the phase estimate $\ket{\omega_k}$, $z_k$ being the probability amplitude and $k$ number of terms as:
    \begin{equation*}
        \ket{\zeta} = \sum_{k=1}^{\lfloor\frac{N}{2}\rfloor} (z_k) \Big[\ket{\omega_k} \otimes \ket{\psi_k} \Big].
    \end{equation*}
    From the density matrix, $\zeta = \ket{\zeta}\bra{\zeta}$, we then trace out the eigenstate $\ket{\psi_k}$ and achieve the state $\Lambda$ as:
    \begin{equation*}
    \Lambda = \sum_{k=1}^{\lfloor\frac{N}{2}\rfloor}(z_k^2)\ket{\omega_k}\bra{\omega_k},
    \end{equation*}
    where $z_k^2$ is the associated probability.

    \item \label{algo_step 7} We perform the density matrix exponentiation of $n^{\prime}$ copies of the density matrix $\Lambda$ to obtain $W = e^{i\Lambda\tau_{2}}$. Next we perform an IQPE with the eigenstates $\ket{0^{\otimes n_1}}$, where $n_1 = \lceil\log_{_{2}}(dim(\Lambda))\rceil$ and consisting of $m^{\prime}$ qubits in the phase estimate register.

    \item \label{algo_step 8} Upon measurement of the phase estimate register, if the result equals the maximum value, i.e.~$\frac{1}{2\pi}$, there exist no winning strategy for $Bob$ and the output is ``$No$'', else if the phase estimate register gives a value less than $\frac{1}{2\pi}$ then output is ``$Yes$'', there is a possible winning strategy for $Bob$.

\end{enumerate}

\section{Algorithm Complexity}
The following analysis shows that the algorithm can be executed on a quantum computer in polynomial time instead of exponential time. 
\begin{enumerate}
    \item \label{complexity 1} Since the unitary $U$, and the Hamiltonian $\Phi$ is a sparse matrix ($s = 1$, $s$ is sparsity of the matrix), $U = e^{i\Phi \tau_{\circ}}$ can be implemented effectively in $O(2\log(M)s^2\tau_{\circ}) = O(2\log(N)\tau_{\circ})$ steps \cite{harrow2009quantum, berry2007efficient}, assuming $M = N$, and each state having $2\log N$ qubits. $(N-1)$ copies of $U$ are created in parallel for $V = U^{\otimes l}$. 

    \item \label{complexity 2} The controlled swap operation acts in parallel on the $\lceil \log N \rceil$ qubits of each of the qubit data registers. Since, there are $^{(N-1)}C_2 = ((N-1)(N-2))/2 = O(N^2)$ series of controlled swaps, it gives a complexity of $O(N^2)$ for step~\ref{algo_step 2}. 

    \item \label{complexity 3} In the Fig.~\ref{fig:register-state} of step~\ref{algo_step 3}, i.e., augmenting the $N$ qubit register state to $2(N-1)$ qubit register state (from $\rho$ to $\sigma$), the complexity is independent of $N$. This is because all the $CNOT$ gates are applied in parallel. 

    \item \label{complexity 4} In step~\ref{algo_step 4}, we use the Improved Quantum Phase Estimation (IQPE) from Ref. \cite{harrow2009quantum} in an iterative manner for $(N-1)$ times. This results in the time variable of $\tau_{\circ}= O((N-1)/\epsilon) = O(N/\epsilon)$ in step \ref{algo_step 1} (See Ref. \cite{harrow2009quantum}). Hence, creating $(N-1)$ copies of $U$ has a circuit depth of $O(2\log (N) ((N-1)/\epsilon)) = O(N\log(N)/\epsilon)$. Here, $\epsilon$ is the precision error in the estimation of $U$. 

    \item \label{complexity 5} In step~\ref{algo_step 5}, the complexity is polynomial for the controlled rotation of the ancilla register state as in Ref. \cite{harrow2009quantum}. 

    \item \label{complexity 6} The dimension of the density matrix $\chi$ generated from the effective state of the rotated ancilla qubits is $2^{(N-1)}$. Therefore, the circuit depth of density matrix exponentiation of $\chi$, i.e., $Y=e^{i\chi\tau_{1}}$ in step~\ref{algo_step 6} of the (single) ancilla qubit is $O(\log(2^{(N-1)})\tau_{1}^2/\varepsilon) = O((N-1)\tau_{1}^2/\varepsilon)$. Here $\varepsilon$ is the simulation error in $Y$. The maximum circuit depth required in the creation of eigenstates $\ket{\psi}$ mentioned in step~\ref{algo_step 6} is $O(N-2) = O(N)$. The complexity of IQPE of $Y$ with the generated eigenstates $\ket{\psi}$ is $O((N-1)/(\eta^2\varepsilon))$. $\eta$ is the estimation precision error for the improved quantum phase estimation for which $\tau_{1} = O(1/\eta)$. 

    \item \label{complexity 7} The dimension of $\Lambda$ is $O(N!) \geq O(2^N)$ and hence the circuit depth for the density matrix exponentiation of $\Lambda$ mentioned in step~\ref{algo_step 7} is $O(\log(2^{N})\tau_{2}^{2}/\kappa) = O({N}\tau_{2}^2/\kappa)$. $\kappa$ is the error in determining the desired accuracy in the creation of $W = e^{i\Lambda\tau_2}$. The complexity of IQPE of $W$ with the eigenstate $\ket{0^{\otimes n_1}}$ (mentioned in step~\ref{algo_step 7}) is $O(N/(\xi^2\kappa))$. $\xi$ is the estimation precision error for improved quantum phase estimation, for which $\tau_2 = O(1/\xi)$.

    \item \label{complexity 18} The complexity of measuring the phase estimate mentioned at the end of step~\ref{algo_step 8} of the algorithm can be ignored.
\end{enumerate}
For simplicity without the loss of generality, we consider $\epsilon = \varepsilon = \eta = \kappa = \xi$. The dominant complexity of our algorithm is $O((N-1)/\epsilon^3)$ of step~\ref{algo_step 6}. In the IQPE process of $Y$ with the eigenstates $\ket{\psi}$, we take $\epsilon = O(1/\text{poly}(N))$ with $T=O(N!)$. Here $\epsilon$ or $\epsilon/2$ is an error in trace distance, and it determines the maximum error probability in the simulation or estimation of $Y$. Furthermore, we repeat step~\ref{algo_step 6} to generate each copy of the density matrix $\Lambda$ required to perform DME efficiently in the simulation of $W$. The IQPE procedure applied in step~\ref{algo_step 7}, has a complexity of $O(N/\epsilon^{3})$, and we choose $\epsilon = O(1/\mathrm{poly}(N))$ with $T = 2^{O(N)}$. From step~\ref{algo_step 6} and step~\ref{algo_step 7} the overall complexity of the algorithm is $O((N-1)/\epsilon^3) \times O(N/\epsilon^3) = O(N^2/\epsilon^6)$, which is polyomial in $N$. The algorithm does not need to provide the correct decision all the time, but at least $(2/3)^{rd}$ of the time, to be in BQP.

\section{Conclusion}
In this work, we presented a bounded-error quantum polynomial-time (BQP) algorithm for solving the generalized geography game (GGG) problem, which is $PSPACE$-complete. The overall time complexity of our algorithm is $O(N^{2}/\epsilon^{6})$, which is polynomial for $\epsilon =$ O(1/$\textrm{poly}\,N$), ensuring the overall cumulative probability of error of our algorithm does not exceed 1/3. 
\par

\bibliographystyle{apsrev4-2}
\bibliography{references}

\appendix
\section{Example of the Algorithm for $N=4$}
\label{appendix}

\begin{figure}[H]
  \centering
  \begin{tikzpicture}
    \node[draw, circle, minimum size=1.5cm] (Q0) at (0, 0) {$\ket{0}$};
    \node[draw, circle, minimum size=1.5cm] (Q1) at (6, 0) {$\ket{1}$};
    \node[draw, circle, minimum size=1.5cm] (Q2) at (6, 6) {$\ket{2}$};
    \node[draw, circle, minimum size=1.5cm] (Q3) at (0, 6) {$\ket{3}$};

    \draw[<->, >=latex, very thick] (Q0) -- 
        node[midway, below] {$\phi_{01} = 0.067$}
        node[midway, above] {$\phi_{10} = 0.067$} (Q1);
    \draw[<->, >=latex, very thick] (Q0) -- 
        node[midway, pos=0.25, rotate = 90, xshift= 25, yshift= 5, left] {$\phi_{03} = 0.067$}
        node[midway, pos=0.25, rotate = 90, xshift= 45, yshift= 5, right] {$\phi_{30} = 0.133$} (Q3);
    \draw[<->, >=latex, very thick] (Q0) -- 
        node[pos=0.25, sloped, shift={(-2mm,-2mm)}] {$\phi_{02} = 0.067$}
        node[pos=0.75, sloped, shift={(2mm,2mm)}] {$\phi_{20} = 0.067$} (Q2);
    \draw[<->, >=latex, very thick] (Q1) -- 
        node[pos=0.25, sloped, shift={(2mm,2mm)}] {$\phi_{13} = 0.067$}
        node[pos=0.75, sloped, shift={(-2mm,-2mm)}] {$\phi_{31} = 0.133$} (Q3);
    \draw[<->, >=latex, very thick] (Q1) -- 
        node[midway, pos=0.25, rotate = 270, xshift=-25, yshift=5, right] {$\phi_{12} = 0.067$}
        node[midway, pos=0.40, rotate = 270, xshift=-25, yshift=5, left] {$\phi_{21} = 0.067$} (Q2);
    \draw[<->, >=latex, very thick] (Q2) -- 
        node[midway, above] {$\phi_{23} = 0.067$}
        node[midway, below] {$\phi_{32} = 0.133$} (Q3);
  \end{tikzpicture}
  \caption{Complete graph with quantum states $\ket{0}$ to $\ket{3}$ with the normalised phase factors in the edges.}
\end{figure}
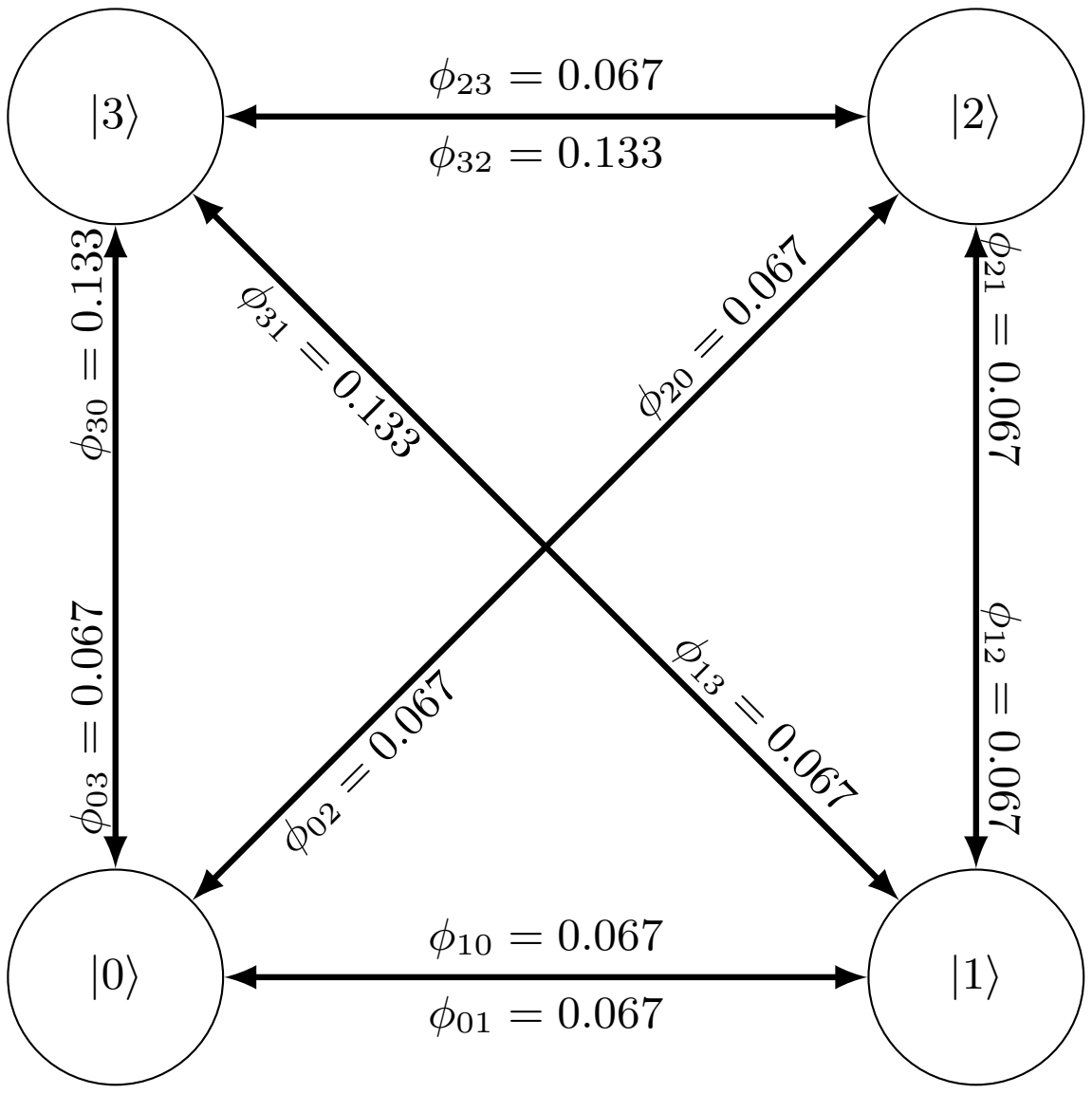

We run the algorithm for a given directed graph with $N = 4$ nodes. The nodes in the graph are $\ket{0} = \text{Agartala}, \ket{1} = \text{Ambala}, \ket{2} = \text{Agra} \hspace{2pt} \& \hspace{2pt} \ket{3} = \text{Assam}$. The phase value of the valid edges $\delta_{01} = \delta_{02} = \delta_{03} = \delta_{10} = \delta_{12} = \delta_{13} = \delta_{21} = \delta_{22} = \delta_{23} = 1$; invalid edges $\delta_{30} = \delta_{31} = \delta_{32} = 2$ and $\delta_{00} = \delta_{11} = \delta_{22} = 0$. For this graph, the total sum of the assigned phase values for $2 \times ^NC_2 = 12$ number of edges is $\phi_{sum} = \sum_{a,b} \delta_{ab}= 15$. Following the step~\ref{algo_step 1} of the algorithm, the normalized phase factors values for the valid edges are $\phi_{01} = \phi_{02} = \phi_{03} = \phi_{10} = \phi_{12} = \phi_{13} = \phi_{20} = \phi_{21} = \phi_{23} = \phi^{V}_{ab} = 0.067$, for the invalid edges are $\phi_{30} = \phi_{31} = \phi_{32} = 0.133$. From the Eq.~\ref{eq : unitrary general}, the unitary $U$ is :
\begin{equation}
\label{eq : U in numbers}
\footnotesize
\resizebox{\columnwidth}{!}{$
U =
\left[
\begin{array}{cccccccccccccccc}
e^{i0} & 0 & 0 & 0 & 0 & 0 & 0 & 0 & 0 & 0 & 0 & 0 & 0 & 0 & 0 & 0 \\
0 & e^{i0.067} & 0 & 0 & 0 & 0 & 0 & 0 & 0 & 0 & 0 & 0 & 0 & 0 & 0 & 0 \\
0 & 0 & e^{i0.067} & 0 & 0 & 0 & 0 & 0 & 0 & 0 & 0 & 0 & 0 & 0 & 0 & 0 \\
0 & 0 & 0 & e^{i0.067} & 0 & 0 & 0 & 0 & 0 & 0 & 0 & 0 & 0 & 0 & 0 & 0 \\
0 & 0 & 0 & 0 & e^{i0.067} & 0 & 0 & 0 & 0 & 0 & 0 & 0 & 0 & 0 & 0 & 0 \\
0 & 0 & 0 & 0 & 0 & e^{i0} & 0 & 0 & 0 & 0 & 0 & 0 & 0 & 0 & 0 & 0 \\
0 & 0 & 0 & 0 & 0 & 0 & e^{i0.067} & 0 & 0 & 0 & 0 & 0 & 0 & 0 & 0 & 0 \\
0 & 0 & 0 & 0 & 0 & 0 & 0 & e^{i0.067} & 0 & 0 & 0 & 0 & 0 & 0 & 0 & 0 \\
0 & 0 & 0 & 0 & 0 & 0 & 0 & 0 & e^{i0.067} & 0 & 0 & 0 & 0 & 0 & 0 & 0 \\
0 & 0 & 0 & 0 & 0 & 0 & 0 & 0 & 0 & e^{i0.067} & 0 & 0 & 0 & 0 & 0 & 0 \\
0 & 0 & 0 & 0 & 0 & 0 & 0 & 0 & 0 & 0 & e^{i0} & 0 & 0 & 0 & 0 & 0 \\
0 & 0 & 0 & 0 & 0 & 0 & 0 & 0 & 0 & 0 & 0 & e^{i0.067} & 0 & 0 & 0 & 0 \\
0 & 0 & 0 & 0 & 0 & 0 & 0 & 0 & 0 & 0 & 0 & 0 & e^{i0.133} & 0 & 0 & 0 \\
0 & 0 & 0 & 0 & 0 & 0 & 0 & 0 & 0 & 0 & 0 & 0 & 0 & e^{i0.133} & 0 & 0 \\
0 & 0 & 0 & 0 & 0 & 0 & 0 & 0 & 0 & 0 & 0 & 0 & 0 & 0 & e^{i0.133} & 0 \\
0 & 0 & 0 & 0 & 0 & 0 & 0 & 0 & 0 & 0 & 0 & 0 & 0 & 0 & 0 & e^{i0}
\end{array}
\right]
$}.
\end{equation}
In our algorithm, $Alice$ (Player $1$) starts the game from ${\ket{0}}$ and there are 6 possible eigenstates with this starting node. Following the step \ref{algo_step 2} and \ref{algo_step 3} of the algorithm we have $\sigma$ from Eq.~\ref{sigma-data-registers} as,
{
\scriptsize
\begin{equation}
\label{eq : sigma in numbers}
\begin{aligned}
\sigma = \frac{1}{8} \Bigg(
\ket{011223}\bra{011223} + 
\ket{011332}\bra{011332} + \\
\ket{022113}\bra{022113} + 
\ket{022331}\bra{022331} + \\
2\ket{033112}\bra{033112} + 
2\ket{033221}\bra{033221}
\Bigg).
\end{aligned}
\end{equation}
}
Now we perform Improved Quantum Phase Estimation (IQPE) of the eigenstates $\sigma$ in  Eq.~\ref{eq : sigma in numbers} with the unitary $U$ in Eq.~\ref{eq : U in numbers} as mentioned in step~\ref{algo_step 4} of the algorithm. In this exmple, we consider perfect phase estimation and therefore, from Eq.~\ref{eq : gamma}, we get $\gamma$ as:

{\scriptsize
\begin{equation}
\label{eq : gamma_numbered}
\resizebox{\columnwidth}{!}{$
\begin{aligned}
\gamma =
\frac{1}{8} \bigg[\ket{0.201}\bra{0.201} \otimes \ket{0.134}\bra{0.134} \otimes \ket{0.067}\bra{0.067}\bigg] \otimes \ket{011223}\bra{011223} \\
+ \frac{1}{8} \bigg[\ket{0.267}\bra{0.267} \otimes \ket{0.134}\bra{0.134} \otimes \ket{0.067}\bra{0.067}\bigg] \otimes \ket{011332}\bra{011332} \\
+ \frac{1}{8} \bigg[\ket{0.201}\bra{0.201} \otimes \ket{0.134}\bra{0.134} \otimes \ket{0.067}\bra{0.067}\bigg] \otimes \ket{022113}\bra{022113} \\
+ \frac{1}{8} \bigg[\ket{0.267}\bra{0.267} \otimes \ket{0.134}\bra{0.134} \otimes \ket{0.067}\bra{0.067}\bigg] \otimes \ket{022331}\bra{022331} \\
+ \frac{2}{8} \bigg[\ket{0.267}\bra{0.267} \otimes \ket{0.200}\bra{0.200} \otimes \ket{0.067}\bra{0.067}\bigg] \otimes \ket{033112}\bra{033112} \\
+ \frac{2}{8} \bigg[\ket{0.267}\bra{0.267} \otimes \ket{0.200}\bra{0.200} \otimes \ket{0.067}\bra{0.067}\bigg] \otimes \ket{033221}\bra{033221}
\end{aligned}
$}
\end{equation}
}
Following the step~\ref{algo_step 5} of the algorithm, for each phase estimate value in Eq.~\ref{eq : gamma_numbered}, we add $n$ ancilla qubits in the state $\ket{0^{\otimes n}}$, giving us $\Omega$ from Eq.~\ref{eq : control-rotation} as, 
{\scriptsize
\begin{equation}
\label{eq : controlled-rotation-numbered}
\resizebox{\columnwidth}{!}{$
\begin{split}
\Omega &= \frac{1}{8}\Bigg[
\Big(
\ket{0.201}\bra{0.201}
\otimes
\ket{0.134}\bra{0.134}
\otimes
\ket{0.067}\bra{0.067}
\Big)
\otimes
\ket{011223}\bra{011223}
\\
&\quad \otimes
\Bigg[
\left(
\sqrt{1-\frac{(0.201)^2}{(0.201)^2}}\ket{0^{\otimes n}}
+
\sqrt{\frac{(0.201)^2}{(0.201)^2}}\ket{1^{\otimes n}}
\right)
\left(
\sqrt{1-\frac{(0.201)^2}{(0.201)^2}}\bra{0^{\otimes n}}
+
\sqrt{\frac{(0.201)^2}{(0.201)^2}}\bra{1^{\otimes n}}
\right)
\\
&\qquad \otimes
\left(
\sqrt{1-\frac{(0.134)^2}{(0.134)^2}}\ket{0^{\otimes n}}
+
\sqrt{\frac{(0.134)^2}{(0.134)^2}}\ket{1^{\otimes n}}
\right)
\left(
\sqrt{1-\frac{(0.134)^2}{(0.134)^2}}\bra{0^{\otimes n}}
+
\sqrt{\frac{(0.134)^2}{(0.134)^2}}\bra{1^{\otimes n}}
\right)
\\
&\qquad \otimes
\left(
\sqrt{1-\frac{(0.067)^2}{(0.067)^2}}\ket{0^{\otimes n}}
+
\sqrt{\frac{(0.067)^2}{(0.067)^2}}\ket{1^{\otimes n}}
\right)
\left(
\sqrt{1-\frac{(0.067)^2}{(0.067)^2}}\bra{0^{\otimes n}}
+
\sqrt{\frac{(0.067)^2}{(0.067)^2}}\bra{1^{\otimes n}}
\right)
\Bigg]
\\[1em]
&\quad +
\Big(
\ket{0.267}\bra{0.267}
\otimes
\ket{0.134}\bra{0.134}
\otimes
\ket{0.067}\bra{0.067}
\Big)
\otimes
\ket{011332}\bra{011332}
\\
&\quad \otimes
\Bigg[
\left(
\sqrt{1-\frac{(0.267)^2}{(0.201)^2}}\ket{0^{\otimes n}}
+
\sqrt{\frac{(0.267)^2}{(0.201)^2}}\ket{1^{\otimes n}}
\right)
\left(
\sqrt{1-\frac{(0.267)^2}{(0.201)^2}}\bra{0^{\otimes n}}
+
\sqrt{\frac{(0.267)^2}{(0.201)^2}}\bra{1^{\otimes n}}
\right)
\\
&\qquad \otimes
\left(
\sqrt{1-\frac{(0.134)^2}{(0.134)^2}}\ket{0^{\otimes n}}
+
\sqrt{\frac{(0.134)^2}{(0.134)^2}}\ket{1^{\otimes n}}
\right)
\left(
\sqrt{1-\frac{(0.134)^2}{(0.134)^2}}\bra{0^{\otimes n}}
+
\sqrt{\frac{(0.134)^2}{(0.134)^2}}\bra{1^{\otimes n}}
\right)
\\
&\qquad \otimes
\left(
\sqrt{1-\frac{(0.067)^2}{(0.067)^2}}\ket{0^{\otimes n}}
+
\sqrt{\frac{(0.067)^2}{(0.067)^2}}\ket{1^{\otimes n}}
\right)
\left(
\sqrt{1-\frac{(0.067)^2}{(0.067)^2}}\bra{0^{\otimes n}}
+
\sqrt{\frac{(0.067)^2}{(0.067)^2}}\bra{1^{\otimes n}}
\right)
\Bigg]
\\[1em]
&\quad +
\Big(
\ket{0.201}\bra{0.201}
\otimes
\ket{0.134}\bra{0.134}
\otimes
\ket{0.067}\bra{0.067}
\Big)
\otimes
\ket{022113}\bra{022113}
\\
&\quad \otimes
\Bigg[
\left(
\sqrt{1-\frac{(0.201)^2}{(0.201)^2}}\ket{0^{\otimes n}}
+
\sqrt{\frac{(0.201)^2}{(0.201)^2}}\ket{1^{\otimes n}}
\right)
\left(
\sqrt{1-\frac{(0.201)^2}{(0.201)^2}}\bra{0^{\otimes n}}
+
\sqrt{\frac{(0.201)^2}{(0.201)^2}}\bra{1^{\otimes n}}
\right)
\\
&\qquad \otimes
\left(
\sqrt{1-\frac{(0.134)^2}{(0.134)^2}}\ket{0^{\otimes n}}
+
\sqrt{\frac{(0.134)^2}{(0.134)^2}}\ket{1^{\otimes n}}
\right)
\left(
\sqrt{1-\frac{(0.134)^2}{(0.134)^2}}\bra{0^{\otimes n}}
+
\sqrt{\frac{(0.134)^2}{(0.134)^2}}\bra{1^{\otimes n}}
\right)
\\
&\qquad \otimes
\left(
\sqrt{1-\frac{(0.067)^2}{(0.067)^2}}\ket{0^{\otimes n}}
+
\sqrt{\frac{(0.067)^2}{(0.067)^2}}\ket{1^{\otimes n}}
\right)
\left(
\sqrt{1-\frac{(0.067)^2}{(0.067)^2}}\bra{0^{\otimes n}}
+
\sqrt{\frac{(0.067)^2}{(0.067)^2}}\bra{1^{\otimes n}}
\right)
\Bigg]
\\[1em]
&\quad +
\Big(
\ket{0.267}\bra{0.267}
\otimes
\ket{0.134}\bra{0.134}
\otimes
\ket{0.067}\bra{0.067}
\Big)
\otimes
\ket{022331}\bra{022331}
\\
&\quad \otimes
\Bigg[
\left(
\sqrt{1-\frac{(0.267)^2}{(0.201)^2}}\ket{0^{\otimes n}}
+
\sqrt{\frac{(0.267)^2}{(0.201)^2}}\ket{1^{\otimes n}}
\right)
\left(
\sqrt{1-\frac{(0.267)^2}{(0.201)^2}}\bra{0^{\otimes n}}
+
\sqrt{\frac{(0.267)^2}{(0.201)^2}}\bra{1^{\otimes n}}
\right)
\\
&\qquad \otimes
\left(
\sqrt{1-\frac{(0.134)^2}{(0.134)^2}}\ket{0^{\otimes n}}
+
\sqrt{\frac{(0.134)^2}{(0.134)^2}}\ket{1^{\otimes n}}
\right)
\left(
\sqrt{1-\frac{(0.134)^2}{(0.134)^2}}\bra{0^{\otimes n}}
+
\sqrt{\frac{(0.134)^2}{(0.134)^2}}\bra{1^{\otimes n}}
\right)
\\
&\qquad \otimes
\left(
\sqrt{1-\frac{(0.067)^2}{(0.067)^2}}\ket{0^{\otimes n}}
+
\sqrt{\frac{(0.067)^2}{(0.067)^2}}\ket{1^{\otimes n}}
\right)
\left(
\sqrt{1-\frac{(0.067)^2}{(0.067)^2}}\bra{0^{\otimes n}}
+
\sqrt{\frac{(0.067)^2}{(0.067)^2}}\bra{1^{\otimes n}}
\right)
\Bigg]
\\[1em]
&\quad +
2\Big(
\ket{0.267}\bra{0.267}
\otimes
\ket{0.200}\bra{0.200}
\otimes
\ket{0.067}\bra{0.067}
\Big)
\otimes
\ket{033112}\bra{033112}
\\
&\quad \otimes
\Bigg[
\left(
\sqrt{1-\frac{(0.267)^2}{(0.201)^2}}\ket{0^{\otimes n}}
+
\sqrt{\frac{(0.267)^2}{(0.201)^2}}\ket{1^{\otimes n}}
\right)
\left(
\sqrt{1-\frac{(0.267)^2}{(0.201)^2}}\bra{0^{\otimes n}}
+
\sqrt{\frac{(0.267)^2}{(0.201)^2}}\bra{1^{\otimes n}}
\right)
\\
&\qquad \otimes
\left(
\sqrt{1-\frac{(0.200)^2}{(0.134)^2}}\ket{0^{\otimes n}}
+
\sqrt{\frac{(0.200)^2}{(0.134)^2}}\ket{1^{\otimes n}}
\right)
\left(
\sqrt{1-\frac{(0.200)^2}{(0.134)^2}}\bra{0^{\otimes n}}
+
\sqrt{\frac{(0.200)^2}{(0.134)^2}}\bra{1^{\otimes n}}
\right)
\\
&\qquad \otimes
\left(
\sqrt{1-\frac{(0.067)^2}{(0.067)^2}}\ket{0^{\otimes n}}
+
\sqrt{\frac{(0.067)^2}{(0.067)^2}}\ket{1^{\otimes n}}
\right)
\left(
\sqrt{1-\frac{(0.067)^2}{(0.067)^2}}\bra{0^{\otimes n}}
+
\sqrt{\frac{(0.067)^2}{(0.067)^2}}\bra{1^{\otimes n}}
\right)
\Bigg]
\\[1em]
&\quad +
2\Big(
\ket{0.267}\bra{0.267}
\otimes
\ket{0.200}\bra{0.200}
\otimes
\ket{0.067}\bra{0.067}
\Big)
\otimes
\ket{033221}\bra{033221}
\\
&\quad \otimes
\Bigg[
\left(
\sqrt{1-\frac{(0.267)^2}{(0.201)^2}}\ket{0^{\otimes n}}
+
\sqrt{\frac{(0.267)^2}{(0.201)^2}}\ket{1^{\otimes n}}
\right)
\left(
\sqrt{1-\frac{(0.267)^2}{(0.201)^2}}\bra{0^{\otimes n}}
+
\sqrt{\frac{(0.267)^2}{(0.201)^2}}\bra{1^{\otimes n}}
\right)
\\
&\qquad \otimes
\left(
\sqrt{1-\frac{(0.200)^2}{(0.134)^2}}\ket{0^{\otimes n}}
+
\sqrt{\frac{(0.200)^2}{(0.134)^2}}\ket{1^{\otimes n}}
\right)
\left(
\sqrt{1-\frac{(0.200)^2}{(0.134)^2}}\bra{0^{\otimes n}}
+
\sqrt{\frac{(0.200)^2}{(0.134)^2}}\bra{1^{\otimes n}}
\right)
\\
&\qquad \otimes
\left(
\sqrt{1-\frac{(0.067)^2}{(0.067)^2}}\ket{0^{\otimes n}}
+
\sqrt{\frac{(0.067)^2}{(0.067)^2}}\ket{1^{\otimes n}}
\right)
\left(
\sqrt{1-\frac{(0.067)^2}{(0.067)^2}}\bra{0^{\otimes n}}
+
\sqrt{\frac{(0.067)^2}{(0.067)^2}}\bra{1^{\otimes n}}
\right)
\Bigg]
\Bigg].
\end{split}
$}
\end{equation}
}
Thereafter, based on the value of the phase estimate in Eq.~\ref{eq : gamma_numbered}, controlled rotation is performed in $\Omega$ in Eq.~\ref{eq : controlled-rotation-numbered}. Subsequently, following step~\ref{algo_step 6} in the algorithm, we trace out the effective state of all the rotated single ancilla qubit and obtain $\chi$ as shown in Eq~\ref{eq : effective_state} as,  
\begin{equation}
\resizebox{\columnwidth}{!}{$
\begin{split}
    \implies \chi = \frac{1}{4} \Bigg[ 2 \Big(\ket{001}\bra{001}\Big)
    + \Big(\ket{011}\bra{011}\Big)
    + \Big(\ket{111}\bra{111}\Big) \Bigg]
\end{split}
$}.
\end{equation}
Afterwards, $\chi$ is exponentiated to obtain the unitary $Y$ as,
\begin{equation}
Y =
\begin{pmatrix}
e^{0} & 0 & 0 & 0 & 0 & 0 & 0 & 0 \\
0 & e^{i0.5} & 0 & 0 & 0 & 0 & 0 & 0 \\
0 & 0 & e^{0} & 0 & 0 & 0 & 0 & 0 \\
0 & 0 & 0 & e^{i0.25} & 0 & 0 & 0 & 0 \\
0 & 0 & 0 & 0 & e^{0} & 0 & 0 & 0 \\
0 & 0 & 0 & 0 & 0 & e^{0} & 0 & 0 \\
0 & 0 & 0 & 0 & 0 & 0 & e^{0} & 0 \\
0 & 0 & 0 & 0 & 0 & 0 & 0 & e^{i0.25}
\end{pmatrix}.
\end{equation}
It undergoes IQPE with eigenstate $\ket{\psi}$ shown in Fig~\ref{fig : QPE with N = 4 cities with 3 edges}a,
\begin{equation}
\ket{\psi} = \frac{1}{\sqrt{2}}(\ket{001} + \ket{111}).
\end{equation}
Subsequently, it gives the output $\ket{\zeta}$ as shown in Eq~\ref{eq : zeta}. In the decimal notation, it is written as,
\begin{equation}
\label{eq : zeta-numbered-binary}
\ket{\zeta} = \frac{1}{\sqrt{2}} \Big[\ket{0.5} \otimes \ket{001} \Big] + \frac{1}{\sqrt{2}} \Big[\ket{0.25} \otimes \ket{111} \Big].
\end{equation}
Continuing with step~\ref{algo_step 6}, from the density matrix of $\zeta = \ket{\zeta}\bra{\zeta}$, we trace out the eigenstates $\ket{\psi}$ thereupon attaining $\Lambda$ as mentioned in Eq.~\ref{lambda} as,  
\begin{equation}
\label{eq : lambda-numbered-decimal}
    \Lambda = \frac{1}{2}\ket{0.5}\bra{0.5} + \frac{1}{2}\ket{0.25}\bra{0.25}.
\end{equation}
Rewriting Eq.~\ref{eq : lambda-numbered-decimal} in binary notation, we have
\begin{align}
\label{eq : lambda-numbered-binary}
\Lambda &= \frac{1}{2}\ket{10}\bra{10} + \frac{1}{2}\ket{01}\bra{01}.
\end{align}
Following step~\ref{algo_step 7}, we exponentiate $\Lambda$ in Eq.~\ref{eq : lambda-numbered-binary} to obtain $W$ as,
\begin{equation}
W = 
\begin{bmatrix}
e^{0} & 0 & 0 & 0\\
0 & e^{\frac{1}{2}} & 0 & 0\\
0 & 0 & e^{\frac{1}{2}} & 0\\
0 & 0 & 0 & e^{0}
\end{bmatrix},
\end{equation}
and perform IQPE with eigenstate $\ket{00}$. The output in the phase estimate register is $0$, which is less than $\frac{1}{2\pi}$. Hence, $Bob$ (Player 2) has a possible winning strategy from the set of eigenstates starting with $\ket{0}$ on this given graph.

\end{document}